%% file: main.tex
\documentclass[sigconf, nonacm]{acmart}

\usepackage{multirow}
\usepackage{diagbox}

\begin{document}
\title{Evaluating NoSQL Databases for OLAP Workloads: A Benchmarking Study of MongoDB, Redis, Kudu and ArangoDB}

%%
%% The "author" command and its associated commands are used to define the authors and their affiliations.
\author{Rishi Kesav Mohan}
\affiliation{%
}
\email{rkmohan2@illinois.edu}

\author{Risheek Rakshit Sukumar Kanmani}
\affiliation{%
}
\email{rrs7@illinois.edu}

\author{Krishna Anandan Ganesan}
\affiliation{%
}
\email{kag8@illinois.edu}

\author{Nisha Ramasubramanian}
\affiliation{%
}
\email{nr50@illinois.edu}
\begin{abstract}
% Fill in this section with your topic sentence outline. Delete or comment out Section 2 before submitting the project proposal. References can be edited in the \texttt{ref.bib} file.

In the era of big data, conventional RDBMS models have become impractical for handling colossal workloads. Consequently, NoSQL databases have emerged as the preferred storage solutions for executing processing-intensive Online Analytical Processing (OLAP) tasks. Within the realm of NoSQL databases, various classifications exist based on their data storage mechanisms, making it challenging to select the most suitable one for a given OLAP workload. While each NoSQL database boasts distinct advantages, inherent scalability, adaptability to diverse data formats, and high data availability are universally recognized benefits crucial for managing OLAP workloads effectively. Existing research predominantly evaluates individual databases within custom data pipeline setups, lacking a standardized approach for comparative analysis across different databases to identify the optimal data pipeline for OLAP workloads. In this paper, we present our experimental insights into how various NoSQL databases handle OLAP workloads within a standardized data processing pipeline. Our experimental pipeline comprises Apache Spark for large-scale transformations, data cleansing, and schema normalization, diverse NoSQL databases as data stores, and a Business Intelligence tool for data analysis and visualization.

The wide-ranging classifications of NoSQL databases include document-oriented, key-value stores, columnar databases, and graph databases. For our experiments, we selected MongoDB, Redis, Apache Kudu, and ArangoDB, each representing a distinct family of NoSQL databases. Leveraging a standardized pipeline, we assessed the performance of these databases using Koalabench, a popular NoSQL benchmarking dataset collection. Each pipeline we setup has a unique NoSQL database and the performance of each pipeline has been evaluated for data loading time and query execution time. We have also compared the performance of a standard SQL solution (PostgreSQL) against the different NoSQL alternatives. Koalabench generated datasets of varying sizes in our desired data model and we conducted a series of experiments using data belonging to two different data models - flat and snow. The insights gleaned from these experiments will facilitate the establishment of an optimal OLAP data pipeline, pairing the ideal NoSQL database as the data warehousing solution.
% \minihead{Problem} 

% \minihead{Set up the Bit}

% \minihead{Introduce the Flip}

% \minihead{Instantiate the flip in a solution}

% \minihead{Evaluation Plan}

% \minihead{Implications}

\end{abstract}

\maketitle

\input{intro}

\input{instructions}
\nocite{*}

%\clearpage
\bibliographystyle{ACM-Reference-Format} % Adjust this to your preferred style
\bibliography{main}
\end{document}

%% file: intro.tex
\section{Introduction}
Modern day data engineering has seen numerous enhancements. Every last ounce of data is being considered as a feature. With data having immense significance in the modern world, the way in which we handle and process data should also evolve with requirements and time. To develop a holistic data engineering pipeline, there are three crucial components:

\begin{itemize}
    \item Data Loading and Processing
    \item Data Store
    \item Data Analysis
\end{itemize}

\subsection{Data Loading and Processing}

Data loading and processing consists of the seamless ingestion, transformation, and preparation of data sourced from various accessible outlets. Given the enormous volume of data to be managed, it requires underlying systems with enhanced processing capabilities to facilitate efficient handling. Apache's Hadoop Distributed File System (HDFS) \cite{hdfs} emerges as one of the prominent solutions built for large-scale data storage. HDFS offers high-throughput access to application data and is ideal for batch processing operations.

With HDFS taking care of large storage, we would have to implement a solution to handle data processing.Data processing is a key aspect of any data pipeline and it governs the process of converting the raw data into a database compatible format before loading it into the database. Apache Spark \cite{spark} stands out as a prime choice for swift and efficient data processing. Its seamless integration with HDFS enables the seamless transfer of data from HDFS to Spark for processing or streaming to subsequent components within the pipeline. Employing these two tools in tandem establishes an optimal platform for transforming data, rendering it ready for import into the designated data store.

\subsection{Data Store}

Once the voluminous data has been stored and processed, we would require a solution for storing the processed data. Ideally, this component would be a database, which helps in storing data in a structured format and it can be accessed using query languages specific to the chosen database. The traditional relational database management system (RDBMS) \cite{rdbms} can store structured data effectively based on a predefined schema. RDBMS also requires complex query operations such as joins in the event of requiring data from more than one table and these complex query operations are generally resource intensive.

With the rise of semi-structured data, there is scope for exploring the possibility of OLAP workloads. NoSQL databases have been adept at leveraging semi-structured data formats such as CSV and JSON which has allowed for the incorporation of a flexible data schema, ease of scalability and support for heterogenous datatypes. NoSQL databases promote flexibility in schema and their data storage mechanism allows for easy access of different data features without the need for executing complex joins to derive insights. 

Despite the benefits of NoSQL over SQL, prevailing architectures have predominantly relied on SQL databases as their primary data store. There exists a significant disparity in evaluating the performance of NoSQL databases within an OLAP-intensive pipeline.
\section{Related Work}
\label{sec:rw}

The literature survey undertaken centered on evaluating prior research within the realms of data processing and databases. These domains were identified as primary areas of interest for assessing the overarching structure of our envisioned pipeline.

\subsection{HDFS and Spark for OLAP}
On the the topic of data processing, we looked far and wide to understand the depth of research that had been performed in order to gauge the progress from the perspective of data engineering pipelines. One of the key ideas, HDFS \cite{hdfspaper} has paved the way for storing large datasets and also provides an invaluable framework for carrying out analysis and transformation operations on the large data using the MapReduce \cite{mapreduce} algorithm. HDFS has paved the way for several database and data streaming solutions to be built on top it. One such data streaming service that has become widely popular is Apache Spark \cite{sparkpaper}. Apache Spark helps in the streaming and processing of voluminous data from HDFS. HDFS and Spark in tandem can handle the load of data processing on any industry-level data engineering pipeline.

Research done based on leveraging HDFS and Spark for OLAP workloads is rather insignificant in the context of NoSQL databases. The standalone research of OLAP pipelines was limited in the previous years, however certain hybrid ideas were proposed. One such idea called Hybrid Transaction/Analytical Processing (HTAP) \cite{htap} proposed for a unified storage for both OLTP and OLAP workloads. While this may seem feasible considering the flexibility of use that systems like HDFS and Spark provide in connecting with different database solutions, this idea does not account for the data rigidity that exists in transactional and analytical processing. RDBMS solutions require a rigid schema and lack support for heterogeneous datatypes whereas NoSQL addresses both concerns. Hence, having a common store and unifying OLTP and OLAP is a little far-fetched for now. 

Wildfire by IBM \cite{wildfire} leveraged a HTAP like solution using HDFS and Spark to perform data analysis and was one of the first solutions to implement a data processing component in their pipeline. This work had stressed on the importance of having distributed data store and distributed data processing to make OLAP more effective. However, Wildfire was based on Spark SQL which is more similar to RDBMS. Stream processing \cite{streamprocessing} became a popular research interest post HDFS and Spark as many ideas revolved around bringing together Spark and OLAP through the metaphorical 'cube'. This process involved transforming raw data using Spark into an OLAP friendly format via a process called cubification. The resultant OLAP cube, contained the features extracted from the provided data, in the form of a cube on top of which data analysis could be performed to gather insights. When distributed processing is employed, the use of OLAP cubes becomes restricted to gathering insights from the processed data and does not aid in determining a scalable strategy for processed data ingestion into the database.

\subsection{Benchmarking for NoSQL-OLAP}
The second broad topic we researched on is the availability of literature for NoSQL databases being used in OLAP/OLTP workloads. This exploration was an effort to understand whether there was a possibility to use NoSQL databases for OLAP. Based on the NoSQL OLAP literature we identified, we also looked for previous benchmarking experiments performed for NoSQL OLAP works. One of the premier works in the NoSQL-OLAP domain involved the creation of OLAP cubes using distributed processing without the need of RDBMS \cite{olapcube}. Using the MC-Cube operator developed in this paper, the idea is to perform transformations on the data stored in columnar NoSQL databases and the data features get bundled into "cubes" which can then be used for analytical insights. This idea came up short because of the use of just a columnar NoSQL database as other NoSQL databases were not tested in their experiments. Another idea on the contrary \cite{nosqlhash} used a document-oriented NoSQL database as the data store solution but employed a series of techniques such as shingling, chunck, minhashing, and locality-sensitive hashing MapReduce. These techniques do serve the purpose of extracting features from the available data but fall short when the data scales. 

Whilst the initial literature stresses more on the usage of NoSQL databases for OLAP workloads, there has been some research done on benchmarking NoSQL databases purely from a query processing perspective for OLAP. The current evaluations of NoSQL databases have primarily focused on simplistic metrics, such as the loading time when used as a data warehouse. However, these comparisons fail to capture the nuanced performance differences within a processing pipeline. Extensive research \cite{olapfornosql,inproceedings} has delved into NoSQL benchmarks, yet these studies often prove inadequate for comprehensive NoSQL database comparisons in a processing context. They emphasize the stark disparities between new benchmarks and traditional Relational benchmarks, illustrating the superiority of the former. For instance, Chevalier et al.'s \cite{olapfornosql} introduction of KoalaBench — a versatile benchmarking tool—enables the analysis of both relational and NoSQL databases, thereby addressing this gap. Numerous studies explore different aspects of NoSQL databases or compare various NoSQL solutions. For instance, \cite{dominguez2010discussion} highlights the significance of Graph Databases in data analysis, while \cite{inproceedings} provides a comparative analysis of three popular NoSQL databases in the context of Big Data and Cloud Computing. Moreover, \cite{10.1007/978-3-319-11587-0_26} details the evaluation of Columnar stores, particularly focusing on HBase and utilizing the Star schema as a metric. Additionally, \cite{dominguez2010discussion} conducts a comparative study between HBase and MongoDB using a star schema setup, further enriching our understanding of these database systems.

\section{The Generic Pipeline}
\label{sec:gp}

The generic pipeline that we envisioned consists of three broad components - data processing, the data store and the data analytics solution. By feeding data to the data processor, we transform the data into a usable format for the data store. All the pipelines we have created uses a different type of NoSQL database as its data store. Finally, all the pipelines culminate with a data analytics solution that reads off the processed data stored in the database to create analytical insights. The pipeline can be further explained in the following sections:

\subsection{Data Processing}
The data processor section consists of two major components - \textit{Apache HDFS} and \textit{Apache Spark}. All the data files required for the pipeline are first loaded on to HDFS. Thanks to the distributed storage and distributed processing capabilities of using HDFS and Spark in tandem, we were able to load data from HDFS and utilize \textit{PySpark} to load the transformed data into the database. 

The raw data we load into HDFS is provided to us by Koalabench \cite{koalabench}. Koalabench can be broadly defined as a decision support benchmark for Big Data requirements. It is derived from the TPC-H benchmark, the reference benchmark in research and industry for decision support systems. It has been adapted to support Big Data technologies such as NoSQL databases and HDFS. It can generate data in different data formats and in different data models. There are four primary data logical models supported by Koalabench, out of which the two data models that we have used are the following :

\begin{figure}
  \centering
  \includegraphics[width=1\linewidth]{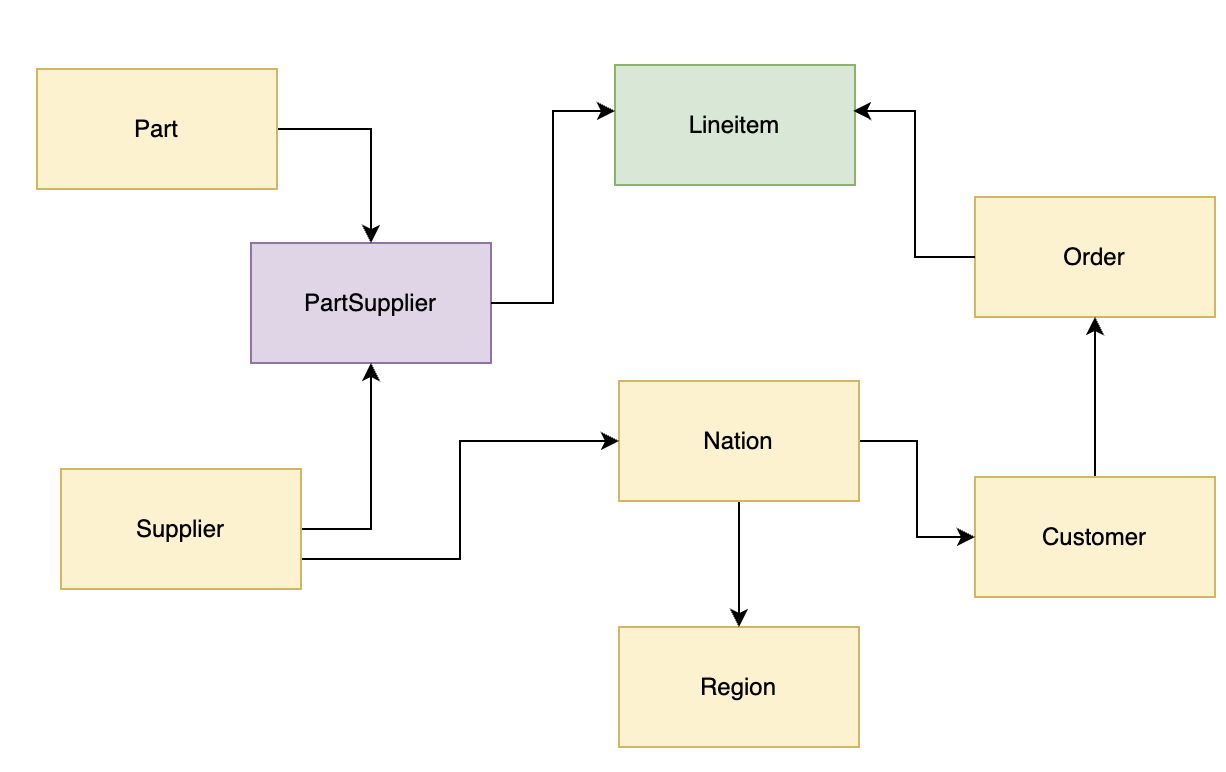}
  \caption{Schema of the dataset based off the \textit{Snow Logical Data Model} generated by Koalabench}
  \label{fig:snowdatamodel}
\end{figure}

\subsubsection{\textbf{Snowflake Data Model}} 
This data model is very close to the one used in the TPC- H benchmark with small modifications.In a classic snowflake data model, there are three major components. \textit{Fact Tables} are the prominent tables based on which most of the data querying is performed - ideally they are the centre of attraction in a snowflake data model. \textit{Component Tables} are supporting tables to the fact table in the sense that they contain most of the required data attributes to know more about the fact table records. \textit{Relationship tables} are used when multiple component tables need to be linked to create a compound component table that aids in providing more information about the fact table records.

As per \autoref{fig:snowdatamodel}, the element in green, Lineitem would be the only fact table for our data model. All tables except PartSupplier will be component tables which aid in providing attributes to Lineitem via simple relationships. PartSupplier is a relationship table composed of attributes from Part and Supplier component tables.

\subsubsection{\textbf{Flat Data Model}} 
\begin{figure}
  \centering
  \includegraphics[width=0.8\linewidth]{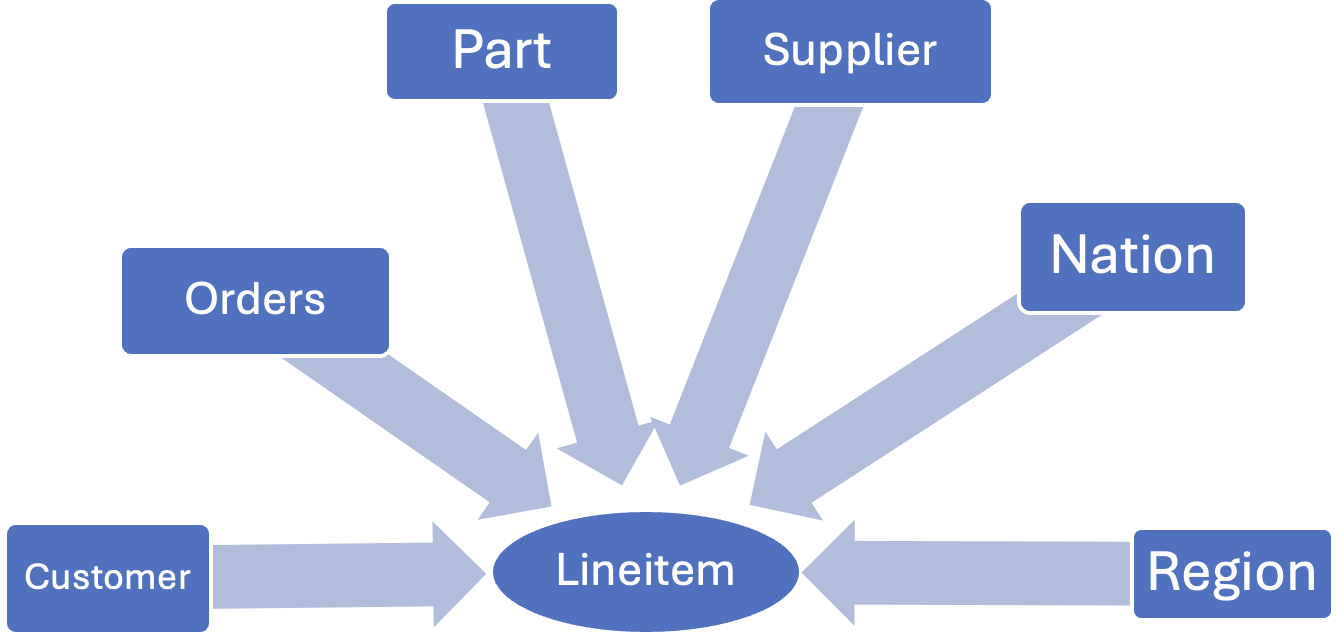}
  \caption{Schema of the dataset based off the \textit{Flat Logical Data Model} generated by Koalabench}
  \label{fig:flatdatamodel}
\end{figure}

This is the simplest data model of all the supported ones. As per \autoref{fig:flatdatamodel}, Lineitem carries over as the premier entity based on which all the data features are developed on. In the flat data model, the significant data parameters from component tables is included within the Lineitem fact table. All relationship tables removed as the flat data model promotes a simpler way of storing data without the need for establishing complex relationships.

The common file format used by all the pipelines for their raw data is CSV and we have utilized the datasets generated using the flat logical data model for the NoSQL databases and the dataset based on the Snowflake logical data model has been used in the PostgreSQL pipeline. This helps us evaluate the performance of the databases with the way in which they handle scalable data upon data ingestion.

The data from HDFS is pushed to Spark for performing data transformations. As our test environment had limited hardware resources to play with, we resorted to utilizing the \textit{num\_partitions} feature of Spark. In Apache Spark, num\_partitions refers to the parameter used to specify the number of partitions to divide an Resilient Distributed Dataset (RDD) or dataframe into during processing. Properly setting num\_partitions can optimize performance by balancing data distribution across the available computing resources. It is essential to choose an appropriate value for num\_partitions to ensure efficient parallel processing and avoid resource contention in Spark applications.

\subsection{Data Store}
Once the processsed data is ready from Spark, we begin the process of inserting the records into the data store component. For simplicity, we have taken up different NoSQL databases to act as the data store for the processed data from Spark. Spark supports interactions with multiple popular SQL and NoSQL databases which has allowed us to use four different NoSQL databases based on the four broad classifications of NoSQL databases. They are as follows:

\subsubsection{\textbf{MongoDB}}
MongoDB \cite{mongosite} is a popular NoSQL database known for its flexible \textit{document-oriented} data model, which stores data in \textit{BSON (Binary JSON)} format. It offers scalability and high performance, with features like sharding and replication for handling large-scale deployments. MongoDB's expressive query language and rich ecosystem of tools make it well-suited for a wide range of applications, from web development to analytics.
\subsubsection{\textbf{ArangoDB}}
ArangoDB \cite{arangosite} is a multi-model database system supporting document, key/value, and graph data models in a single database core. It offers a versatile query language, AQL (ArangoDB Query Language), and boasts features like distributed graph processing and geo-spatial indexing. With its flexible data model and powerful querying capabilities, ArangoDB is suitable for diverse use cases including social networks, recommendation engines, and real-time analytics. For the purposes of our experiments, we have leveraged the graph data models of ArangoDB owing to the \textit{Graph databases} family of NoSQL solutions.
\subsubsection{\textbf{Apache Kudu}}
Apache Kudu \cite{kudusite} is an \textit{open-source columnar storage engine} designed for fast analytics on rapidly changing data. It combines the performance of traditional columnar databases with the ease of use of Hadoop. With features like automatic partitioning and fault tolerance, Apache Kudu is ideal for real-time data ingestion and interactive analytics applications.
\subsubsection{\textbf{Redis}}
Redis \cite{redissite} is an in-memory data store known for its high performance and versatility in caching, session management, and real-time analytics. It supports various data structures like strings, hashes, lists, sets, and sorted sets, making it suitable for a wide range of use cases.

Apart from the mentioned NoSQL databases, we have performed one additional experiment using \textbf{PostgreSQL} \cite{psqlsite} as the RDBMS solution for the data store. This experiment has helped us compare and contrast the performance of SQL and NoSQL databases in the generic pipeline. The processed data has been saved on the respective databases thanks to the versatility provided by \textbf{PySpark} in developing seamless interaction modules between Spark and the databases. 

Once the data has been loaded, we have executed a subset of the queries belonging to the TPC-H benchmark. As Koalabench derives its existence from TPC-H, it has allowed us to reuse the same template of the documented benchmark queries. However, based on how the data is stored in the different databases, we have come up with equivalent queries for each of the NoSQL databases and executed them on the databases respectively.

\section{Architecture}
\label{sec:eval}
For the evaluation of each proposed data pipeline as part of our benchmarking process, we plan to execute datasets of varying scale factors generated from the Koalabench suite. These datasets will be run through each pipeline within our standardized local testing environment, which consists of a system running \textit{MacOS Sonoma 14.4.1}, powered by an \textit{Apple M2 chip} with \textit{8GB of RAM} and \textit{128GB of storage}. To support distributed data processing capabilities for Hadoop and Spark, we intend to create a dockerized environment within the local system. This environment will comprise a master-worker setup facilitating the execution of distributed tasks. 
\begin{figure}
  \centering
  \includegraphics[width=0.5\linewidth]{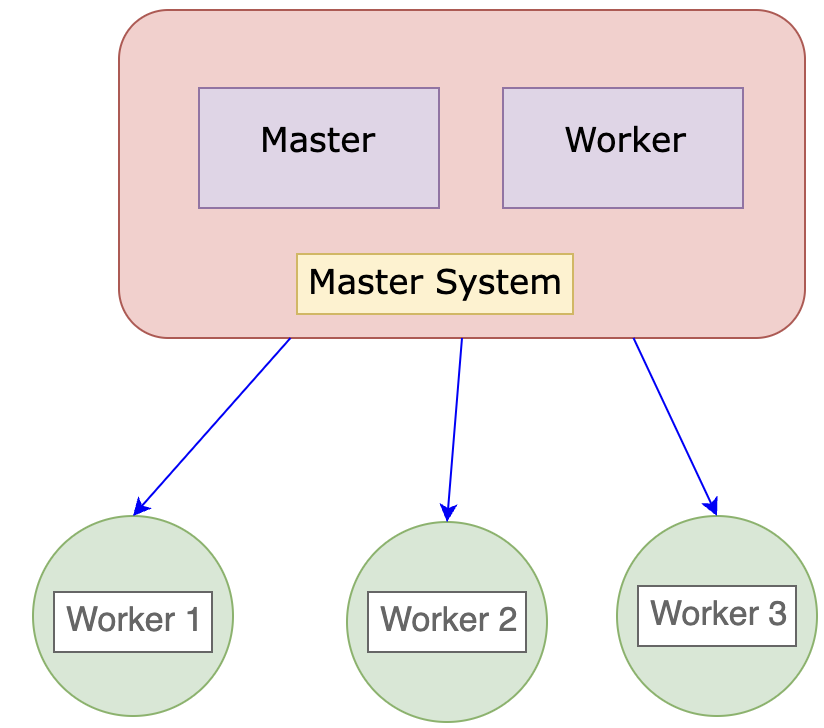}
  \caption{Dockerized experimental setup consisting of one master and three worker nodes}
  \label{fig:architecture}
\end{figure}

The architecture shown in \autoref{fig:architecture} consists of one master node concurrently acting as a worker, along with three dedicated worker nodes. The master node maintains bidirectional communication routes with each worker node, enabling seamless coordination for distributed processing across HDFS and Spark clusters. The dockerized environment is built on \textit{Debian Linux v11} across all nodes. To facilitate distributed data processing capabilities, each node is equipped with \textit{Hadoop-3.3.6} and \textit{Spark-3.4.1}.

We have incorporated the latest docker images for all the databases under evaluation, including \textit{Postgres-alpine-3.19}, \textit{MongoDB-7.0.8}, \textit{ArangoDB-3.12.0}, \textit{Redis-7.2.4}, and \textit{Apache Kudu-1.17}. These database images have been instantiated within the same docker network as the experiment nodes, enabling efficient measurement of time-based metrics without introducing potential biases stemming from network latency. The data analysis component of our pipeline will leverage a Business Intelligence (BI) tool connected to the respective database docker image for accessing and analyzing stored data. The tentative BI tools identified for this purpose are \textit{Metabase v0.49.3} and \textit{Tableau 2024.1}. The BI container will reside within the same docker network as the experimental setup to ensure seamless integration and communication. 

\section{Experiments}
The following section elaborates on the dataset that we have used to evaluate the performance of the various pipelines and documents the findings from the individual experiments we have carried out in each data pipeline we developed.

\subsection{Dataset}
To evaluate our pipelines across varying data scales, we have used different datasets generated from Koalabench using the flat data model based on varying scale factor(sf). The primary objective of utilizing different sf values is to maximize the dataset size, enabling comprehensive testing of our pipelines' performance and scalability across a broad range of data volumes. We have conducted experiments on all the pipelines using sf1, sf2, sf3, sf4 and sf5 datasets with the data format set as CSV for all datasets. The minimum size of the collection belongs to the sf1 dataset at 2.38 GB and the maximum size of the collections belongs to sf5 which scales up to approximately 11GB. For the PostgreSQL pipeline, we have leveraged the snow data model to generate datasets ranging from sf1 to sf5 in the CSV format. The least size is with sf1 at 1.4GB and the maximum size is with sf5 at approximately 6GB.

\subsection{Data Loading Time}

\autoref{tab:loadtime} portrays the load times for different datasets across the five different databases.

\begin{table}[htbp]
\centering
\caption{Data Loading times for all databases across the five datasets (time in seconds)}
\label{tab:loadtime}
\begin{tabular}{|c|c|c|c|c|c|}
\hline
Databases & SF1 & SF2 & SF3 & SF4 & SF5 \\ \hline
PostgreSQL & 37s & 375s & 857s & 1089s & 1481s\\ \hline
MongoDB & 90s & 1250s & 1701s & 2275s & 2810s \\ \hline
ArangoDB & 295s & 2249s & 3964s & 12169s & 15162s \\ \hline
Redis & 1495s & 3245s & 5023s & 7748s & 10289s\\ \hline
Apache Kudu & 42s & 95s & 146s & 192s & 240s\\ \hline
\end{tabular}
\end{table}

\textit{Apache Kudu} has shown the smallest load times across all five datasets. One of the primary reasons for this observation is the columnar-storage mechanism that Apache Kudu employs coupled with its superior in-memory management that flushes out records as and when the threshold is reached. This allows Kudu to handle scaling data efficiently and keep the overall loading time to agreeable levels.

\textit{MongoDB} seems to have progressively higher data ingestion times. MongoDB's document-oriented storage model and lack of native sharding capabilities may lead to increased data ingestion times due to increased indexing and write operations overhead. Additionally, as the dataset size grows, MongoDB's reliance on memory-mapped files for storage may result in slower write performance and increased disk I/O operations.

\textit{Redis} displays the poorest performance on all 5 datasets. The very high data loading times observed in Redis can be attributed primarily to the schema we used for the processed data. In order to keep the schema uniform across all NoSQL databases, we opted for having the \textit{linenumber\_id} as the only key and all the remaining data features were aggregated into a single list and attributed to the corresponding \textit{linenumber\_id}. In general, we ended up creating a record that had a single key and 40 values. Additionally, redis being a single threaded solution meant that for every insertion operation had two call operations to make. With data scaling, this would increase exponentially and in the end be double of what an ideal PostgreSQL implementation would execute.

\textit{ArangoDB} was implemented using its document model and its data insertion times are exorbitant owing to the document model that was implemented. ArangoDB is definitely not an ideal solution should the data scale as for sf5, the data insertion time is almost 50x what it was for sf1.

\textit{PostgreSQL} shows a gradual increase in data insertion times from sf1 all the way upto sf5. It is able to handle scaling data much more efficiently has compared to some of its NoSQL counterparts.

Based on this experiment, it is clear that with a NoSQL database as your data store, it is preferrable to choose a columnar storage solution in case the pipeline has to be robust enough to handle scaling data.

\subsection{Query Processing Time}

As Koalabench is based off the TPC-H benchmark, we selected five out of the seventeen benchmark queries that TPC-H has to offer. The TPC-H benchmark is meant for running benchmark experiments on RDBMS solutions. Hence, the queries can be directly applied to the PostgreSQL data pipeline as the schema is same as the one proposed by TPC-H. For all NoSQL data pipelines that leverage the flat data model, we prepared queries in the native querying language of the NoSQL database present in the pipeline.

The five queries we selected were different from one another based on the significant operation that was being performed by the query. For context, Query 1 was aggregation intensive in which almost eight out of the ten values that were being fetched were based off aggregation. Query 2 had a good balance of aggregation and join operations. Query 3 had an aggregation along with a sub-query based filter. Query 4 had five join operations and Query 5 was the simplest query off them all with just a single aggregation operation.

\begin{figure}
  \centering
  \includegraphics[width=1\linewidth]{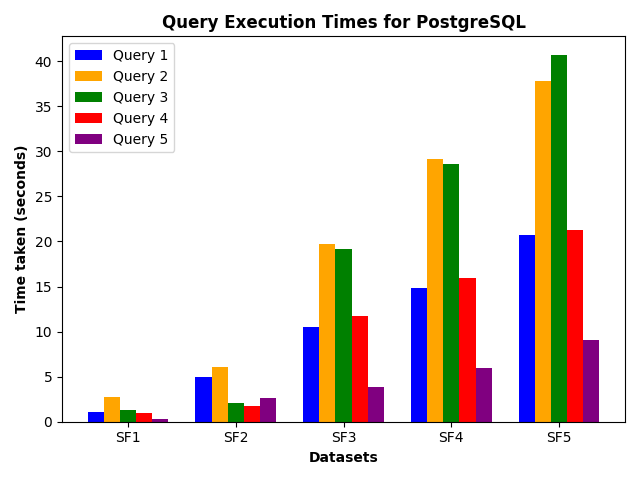}
  \caption{Query Execution time (in seconds) for PostgreSQL across the five datasets}
  \label{fig:psql}
\end{figure}

\subsubsection{\textbf{PostgreSQL}}
\autoref{fig:psql} shows a graph depicting the query execution times for the PostgreSQL data pipeline. Across the five datasets and for the five queries, we see a steady increase by a factor of 2x for almost all queries except Queries 3 and 4. The reason for the observation of exponential rise in query execution times for those two queries alone is because of the inherently costly join operations of RDBMS solutions which tend to compare every row of the two tables. For filtering and aggregation queries, the increase in query execution time is 0.33x for every 2x scale in data which is pretty commendable.

\subsubsection{\textbf{MongoDB}}
\begin{figure}
  \centering
  \includegraphics[width=1\linewidth]{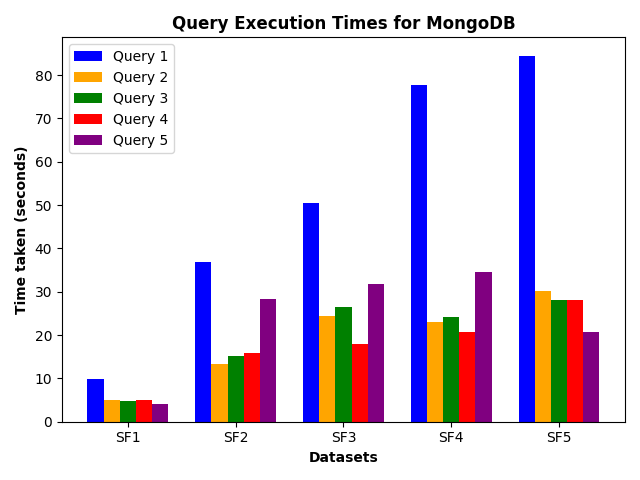}
  \caption{Query Execution time (in seconds) for MongoDB across the five datasets}
  \label{fig:mongo}
\end{figure}
\autoref{fig:mongo} shows a graph displaying the query execution times for the MongoDB pipeline for the five selected benchmark queries across the five datasets. MongoDB uses the translated version of the benchmark queries and the major change between the original and translated versions are that all join related operations become filter related operations. This is because of the non-availability of support for joins in NoSQL databases.

MongoDB handles most of the queries easily with scaling data with a maximum increase in execution times by about 0.8x for every 2x increase in data. This can be attributed to MongoDB's inbuilt sharding capabilities, that promote horizontal scaling approach that allows it to handle larger data sets and provide high throughput operations by distributing data across multiple shards. One aspect where MongoDB struggles is with intense aggregation queries such as Query 1. With data scaling, MongoDB is forced to look up multiple document objects and perform compute operations on them which are CPU intensive and tend to consume more time than rest of the query operations.

\begin{figure}
  \centering
  \includegraphics[width=1\linewidth]{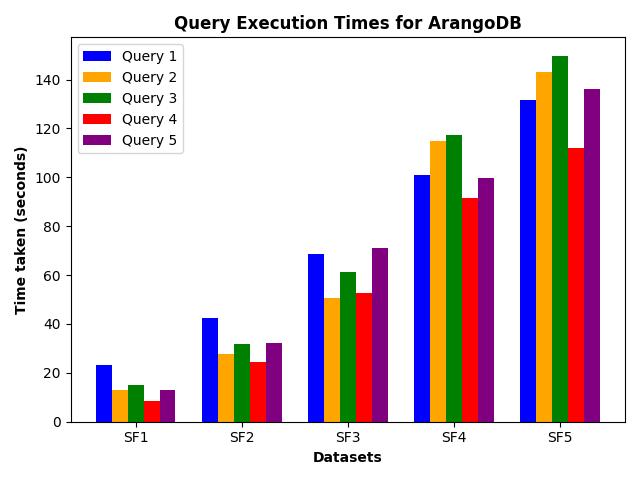}
  \caption{Query Execution time (in seconds) for ArangoDB across the five datasets}
  \label{fig:arango}
\end{figure}

\subsubsection{\textbf{ArangoDB}}
\autoref{fig:arango} shows a graph displaying the query execution times for the ArangoDB pipeline for the five selected benchmark queries across the five datasets. ArangoDB uses the Arango Query Language (AQL) version of the benchmark queries. ArangoDB sees a 1.5x scale in query execution times for every 2x scale in data size. Whilst ArangoDB does not selectively outperform other data pipelines in certain scenarios, it does show a steady increase in query execution time with scaling data. This can be attributed to the dynamic query optimizer that selects the perfect query execution plan depending on the query being executed. ArangoDB employs efficient memory management techniques to minimize disk I/O and maximize query performance. It utilizes memory caches for frequently accessed data and implements buffer management strategies to optimize disk access.

\begin{figure}
  \centering
  \includegraphics[width=1\linewidth]{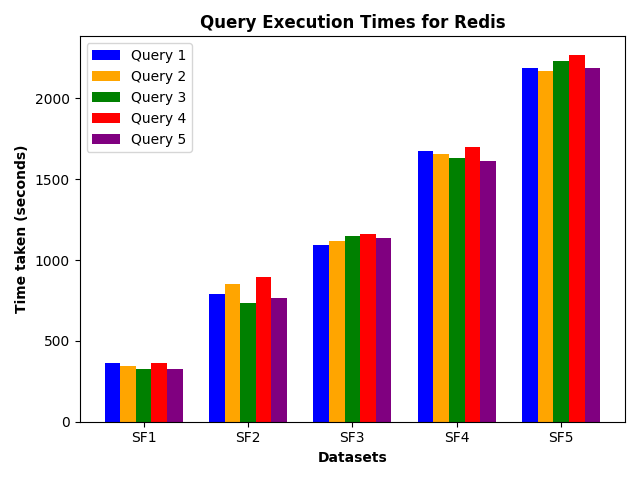}
  \caption{Query Execution time (in seconds) for Redis across the five datasets}
  \label{fig:redis}
\end{figure}

\subsubsection{\textbf{Redis}}
\autoref{fig:redis} portrays a graph containing information on the query execution times of the five benchmark queries across the five datasets on Redis. Irrespective of the size of data, Redis takes longer than all the other databases to execute the queries. For sf1, the smallest dataset, Redis displays a best case of 3x increase in terms of query execution times. Redis suffers primarily from the data schema we had selected for the flat data model. By assigning all features to a single key, \textit{linenumber\_id}, a single row of record becomes huge for redis to retrieve from memory. By default, redis is an in-memory intensive data store and hence, holding such a big record in memory is bound to make the query execution slow and CPU intensive.

\begin{figure}
  \centering
  \includegraphics[width=1\linewidth]{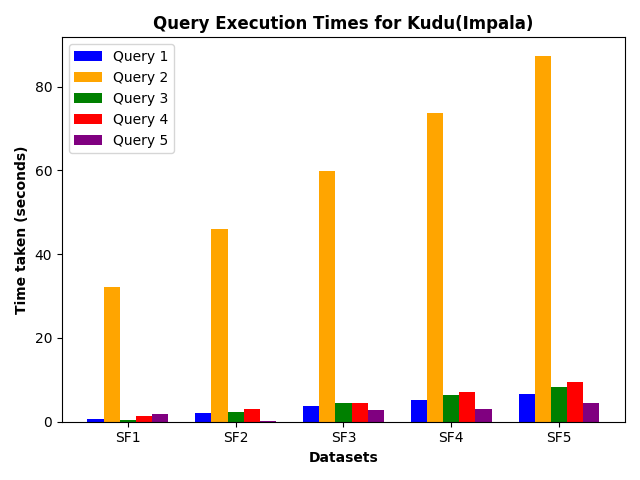}
  \caption{Query Execution time (in seconds) for Apache Kudu across the five datasets}
  \label{fig:kudu}
\end{figure}

\subsubsection{\textbf{Apache Kudu}}
\autoref{fig:kudu} shows the recorded query execution times for the five benchmark queries across the five datasets. Based on these readings, it is clear that Apache Kudu performs the best amongst all the selected databases in terms of query execution. Kudu stores data in a columnar format and this is highly efficient for analytical workloads because it allows queries to read only the columns needed for the query, minimizing I/O operations and improving query performance. This can be visibly seen from the very minimal increase in query execution times across datasets for queries 2 to 5. Even when data scales at 2x, query execution times have tended to stay flat and do not see a proportional or exponential increase like the rest of the databases.
The reason why query 2 sees an exponential increase with scaling data is related to a potentially incorrect time measurement on behalf of Apache Kudu. When fetching data records based on given filter conditions, Kudu tends to write the retrieved records onto the impala shell. The write operation time is also included in the overall execution time of the query and thus it seems beefed up with scaling data compared to the rest of the queires.

\section{Conclusion}

Through this paper, we envisioned our idea of building the ideal data engineering pipeline, that would be take care of data processing and data store. We experimented with datasets of varying scale from the Koalabench dataset where data was generated in two different data models. Our experiments measured the query execution times for 5 benchmark queries and recorded the data ingestion time for all databases. The effect of the data schema was evident in the varied data ingestion times. Certain databases did not handle the scaling levels of data and ended up seeing exponential increase in the query execution times of complex queries. A definitive finding from our experiments is the prominence of Columnar data storage options in future OLAP research. Columnar storages tend to load data 2x faster than other SQL and NoSQL options and the same goes for query execution as well. 

These experiments also highlight the immense scaling and distributed processing capabilities that modern-day NoSQL solutions have and how these can be put to use to solve complex data analytical problems. We believe this idea would be a pioneer towards the adoption of more NoSQL-based data store solutions for evaluating OLAP workloads.
As part of future work, we do recommend the integration of a data analysis tool to commensurate the data pipeline. Research can be extended across various columnar storage options to identify the ideal columnar storage option to evaluate OLAP workloads.
\label{sec:end}